# VISIBLE LIGHT COMMUNICATION USING LED-BASED AR MARKERS FOR ROBOT LOCALIZATION

Wataru Uemura[1] and Shogo Kawasaki[2]

[1]Faculty of Advanced Science and Technology, Ryukoku University, Shiga, Japan
[2]Faculty of Science and Technology, Ryukoku University, Shiga, Japan

## ABSTRACT

*A method of information transmission using visual markers has been widely studied. In this approach, information or identifiers (IDs) are encoded in the black-and-white pattern of each marker. By analyzing the geometric properties of the marker frame - such as its size, distortion, and coordinates - the relative position and orientation between the camera and the marker can be estimated. Furthermore, by associating the positional information of each marker with its corresponding ID, the position of the camera that takes the image picture can be calculated. In the field of mobile robotics, such markers are commonly utilized for robot localization.*

*As mobile robots become more widely used in everyday environments, such visual markers are expected to be utilized across various contexts. In environments where robots collaborate with humans - such as in cell-based manufacturing systems in factories or in domestic settings with partner robots - it is desirable for such markers to be designed in a manner that appears natural and unobtrusive to humans.*

*In this paper, we propose a method for implementing an ArUco marker in the form of illumination. In the proposed method, LEDs are arranged in accordance with the grid pattern of the marker, and the blinking frequency of each LED is determined based on the corresponding black or white cell. As a result, the illumination appears uniformly bright to the human eye, while the camera can capture variations in the blinking frequency. From these differences, the black-and-white pattern can be reconstructed, enabling the identification of the marker's tag information.*

*We develop a prototype system, and conduct experiments which are conducted to evaluate its performance in terms of recognition accuracy under varying distances and viewing angles with respect to the ArUco marker.*

## 1. INTRODUCTION

In many developed countries, including Japan, the declining birthrate and aging population have led to a serious labor shortage. This issue is particularly severe in Japan, where replacing human labor with autonomous mobile robots has become an urgent challenge. As a result, societies in which humans and robots work together in the same environment are beginning to emerge.

For such mobile robots, self-localization is important for autonomous navigation. While outdoor localization is widely achieved using the Global Positioning System (GPS) [1], various methods have been proposed for indoor environments where GPS signals are unavailable. This paper focuses on self-localization using AR markers [2] which is one of the indoor self-localization methods. In AR marker–based localization, markers with unique ID [3, 4] must be placed in the environment. In automated factory, this can be accomplished by attaching printed markers on paper throughout the workspace [5, 6]. However, when considering daily living environments, such markers may appear unnatural or intrusive to humans. Therefore, the markers must blend seamlessly into everyday objects.

To address this issue, we propose constructing AR markers using LED lights that function as ordinary room illumination. In this approach, the lighting appears normal and comfortable to humans, while it simultaneously serves as a marker recognizable by a robot's camera. By embedding marker information in this manner, the robot can obtain ID data and estimate its own position based on the marker's geometric configuration. This approach enables humans and robots to coexist in the same space without altering the familiar appearance of the environment.

The remainder of this paper is organized as follows. Section 2 provides an overview of self-localization methods for mobile robots, including odometry, GPS-based localization, visible light positioning, and AR marker–based localization. Section 3 describes the characteristics of digital cameras, with a focus on the rolling shutter effect and its impact on imaging fast-blinking light sources. Section 4 presents the proposed method, including the design of LED-based AR markers and the implementation of marker-shaped lighting for information transmission in living environments. Section 5 details the experiments conducted to evaluate the recognition performance of the proposed method under varying distances and viewing angles and discusses the results. Finally, Section 6 concludes the paper, summarizing the main findings and discussing future directions for improving marker recognition and system implementation.

## 2. SELF - LOCALIZATION METHODS

When a mobile robot runs autonomously, as illustrated in Figure 1, both the robot's current position and the route information to its destination are essential. This paper focuses on estimating the robot's own position.

Typically, a robot estimates its position by calculating its traveled distance based on the motion of its wheels, a process known as odometry. In this approach, accurate information about the wheel size and placement is critical; any deviation in these parameters causes the estimated position to drift over time as the robot moves. To compensate for such errors, various methods have been developed that use environmental information to improve self-localization accuracy.

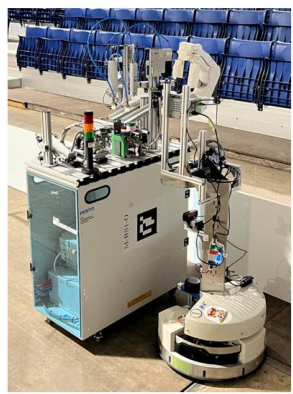

Figure 1. A mobile robot requires the current position to move to the destination using AR marker.

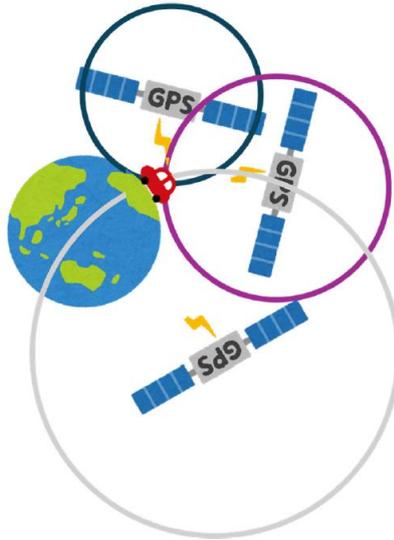

Figure 2. GPS can estimate the receiver's position using three or more satellites.

For outdoor environments, the Global Positioning System (GPS) [1] is widely used, for example, in car navigation systems and autonomous vehicles. In contrast, for indoor environments or areas where satellite signals are unavailable, self-localization can be achieved by using onboard sensors together with pre-mapped environmental data. Another approach employs visual markers, such as two-dimensional barcodes [3, 4], where the relative position and orientation can be computed from the marker's size, inclination, and appearance in the captured image.

This section provides an overview of these self-localization techniques.

## 2.1. Global Positioning System (GPS)

The Global Positioning System is a satellite-based positioning method, as illustrated in Figure 2. In this system, radio signals transmitted from multiple satellites are received by a ground-based receiver to calculate the receiver's position coordinates. Each satellite transmits signals containing precise time information. By measuring the difference between the transmission time from the satellite and the reception time at the receiver, the distance between them can be calculated. With signals from three satellites, a three-dimensional position can be calculated geometrically, but a fourth satellite is required to correct the receiver's internal clock error and obtain an accurate position fix.

When the satellite signals are clearly received - such as in open outdoor environments - the positioning accuracy is high. However, in locations where satellite signals are blocked, such as inside tunnels, or where strong reflections occur, such as in urban canyons between tall buildings, GPS-based self-localization becomes unreliable or unavailable.

## 2.2. Visible Light Positioning

Visible light positioning (VLP) is a method that utilizes visible light communication (VLC) [7, 8] technology. As illustrated in Figure 3, information is embedded in the rapid blinking patterns of light sources, such as LEDs, which are imperceptible to the human eye [9]. The position of a receiver can then be estimated by capturing these signals using a camera or other image sensor [10 – 13].

International Journal of Mobile Network Communications & Telematics (IJMNCT)
Vol.15, No.6, December 2025

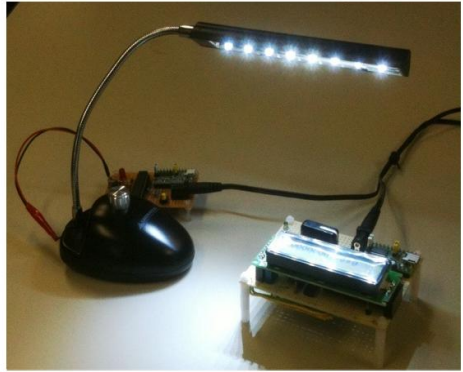

Figure 2. GPS can estimate the receiver's position using three or more satellites.

Care must be taken in the way information is transmitted to ensure that the ratio of on-time to off-time remains constant. If this ratio varies with the transmitted information, the lighting may appear to flicker to humans [14, 15].

While indoor lighting must be replaced with dedicated VLP-enabled luminaires, most indoor environments already use lighting, so installation is relatively straightforward. However, when using a camera as the receiver, a high-performance camera with a high frame rate is typically required, and fisheye lenses may be necessary to capture the entire room, which increases system cost.

## 2.3. AR Marker–based Localization

Two-dimensional barcodes, commonly referred to as markers, can be used for camera pose estimation. Square markers, such as those illustrated in Figure 1, consist of a simple pattern of cells. Among them, ArUco markers [2], which employ a grid-based binary pattern, are readily available through the OpenCV library [16] and can be easily utilized. In this paper, we use ArUco markers to acquire ID information for self-localization.

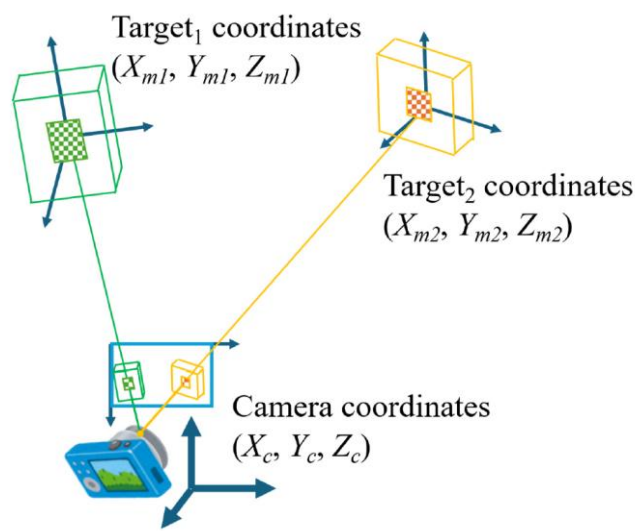

Figure 4. Relationship between camera and targets' coordinate systems.

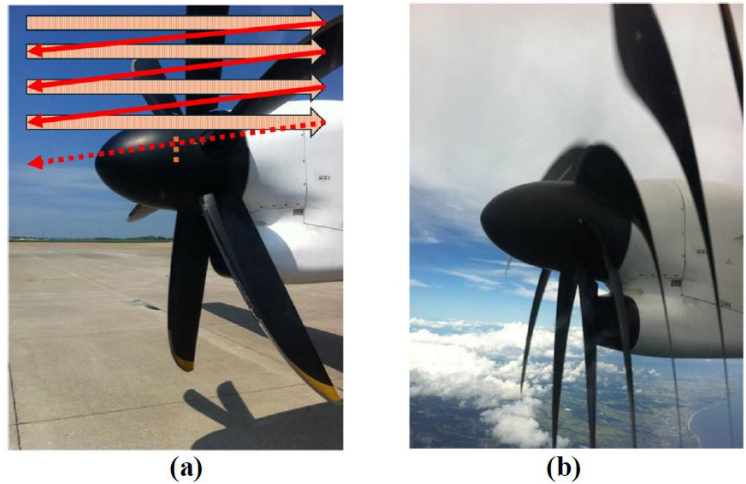

(a)          (b)

Figure 5. Example of rolling shutter effect: (a) a stationary propeller with scanning-line arrows, and (b) a distorted rotating propeller captured with a rolling shutter.

The coordinates of each marker are associated with its unique ID, and the camera position is estimated using the relative distance between the camera and the marker detected in the captured image. As illustrated in Figure 4, the apparent position and size of the marker in the image vary with the distance to the camera. Therefore, camera calibration is performed in advance using a checkerboard pattern to determine the camera's intrinsic parameters. When the marker is rotated relative to the camera, it appears distorted in the image, and this distortion is used to estimate the camera's pose. By combining the estimated camera-to-marker pose with the marker's pre-associated coordinates, the camera's position can be determined.

It should be noted that the spatial coordinates of the markers must be prepared in advance. Moreover, because AR markers must be physically placed in the environment, visual interference with the existing scenery may occur.

## 3. ROLLING SHUTTER EFFECT ON DIGITAL CAMERA

Some digital cameras do not capture the entire image at once but rather scan the scene line by line to acquire image data. When the scene changes during the scanning process, these changes appear in the captured image, as illustrated in Figure 5. For example, when imaging a rapidly blinking light source, alternating bright and dark bands appear along the scanning direction due to the blinking.

The width $w$ of the light band caused by the rolling shutter effect during high-speed blinking can be determined in terms of the camera exposure time $T$, the vertical resolution $n$, and the blinking frequency $f$ (shown in Figure 6). This relationship is expressed by the following equation:

$$w = \frac{n}{T} \times \frac{1}{2f}$$

This equation shows that the band width decreases as the blinking frequency increases or the exposure time decreases.

21

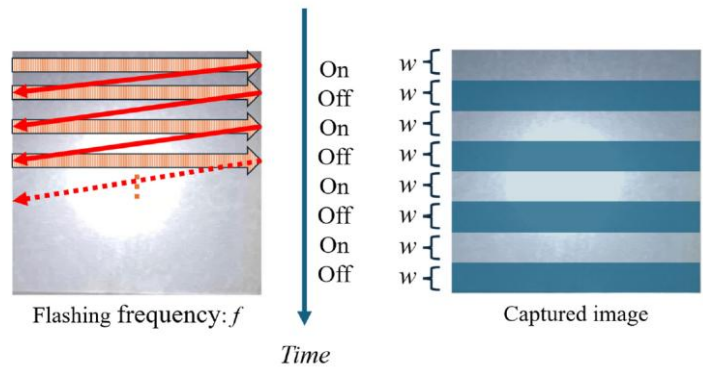

Figure 6. The rolling shutter effect causes the blinking LEDs to appear as alternating on–off light bands in the captured image.

## 4. PROPOSED METHOD

### 4.1. LED-based AR Marker (Marker-shaped Lighting)

We propose an information transmission method using AR markers that does not interfere with the living environment. This is achieved by creating AR markers with multiple light sources and recognizing them through image processing to receive the marker ID information.

### 4.2. Detection Algorithm

The overall workflow of the program is illustrated in Figure 7. The marker created using high-speed blinking LEDs is shown in (a). Two different blinking frequencies are used. Although differences in brightness and individual light sources can be clearly observed in the camera image, these differences are imperceptible to the human eye, and the marker appears simply as a uniformly bright light. The image captured by the camera is shown in (b), where the rolling shutter effect can be observed. To clarify the light bands corresponding to each cell, the image is binarized as shown in (c). Based on the band width, the type of each cell can be classified, and in this example, the two types are distinguished using red and blue colors. The result is shown in (d).

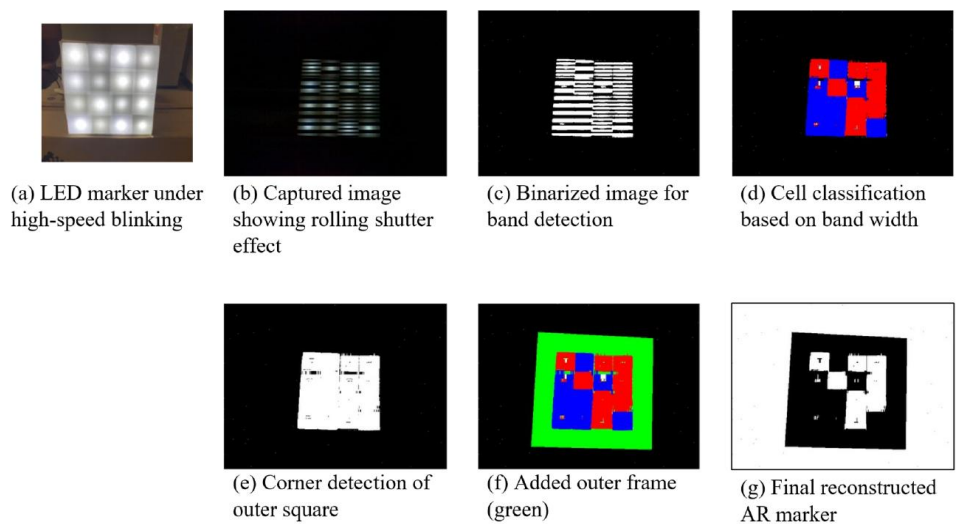

Figure 7. Detection algorithm of the proposed method.

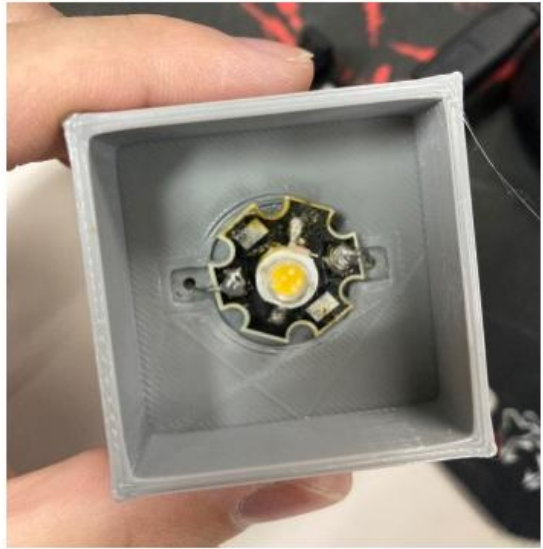

Figure 8. One LED is located in one frame.

Next, since a border is required for marker recognition, the image is binarized again so that the entire marker becomes a white square, and the four corners of the square are detected, as shown in (e). From the detected square size, the size of one cell is estimated, and an additional frame equivalent to one cell width is added around the outer edge. The added frame is shown in green in (f). The cell information within the frame is restored from (d), where the red and blue cells are mapped back. Finally, by converting the outer frame to white and the inner red and blue cells to black and white respectively, the marker pattern is completed, as shown in (g).

As shown in Figure 8, the LEDs are arranged in a grid. A 4 × 4 LED grid is used in the prototype (Figure 9 (a)); the proposed method, however, is applicable to grids of any size. The surface is covered with paper as shown in Figure 9 (b) to create the information-bearing areas of the AR marker. Finally, the LEDs are illuminated according to the encoded pattern, as illustrated in Figure 9 (c).

## 5. EXPERIMENTS AND RESULTS

### 5.1. Objective

The objective of the experiments is to evaluate the performance of the proposed AR marker-shaped lighting method by comparing the recognition rate of ArUco markers at various distances and angles with conventional printed markers.

### 5.2. Experimental Conditions

The imaging conditions using the camera (Figure 10) are as follows:

- Resolution: 640 × 480
- Shutter type: rolling shutter
- Exposure time: 0.01 s

23

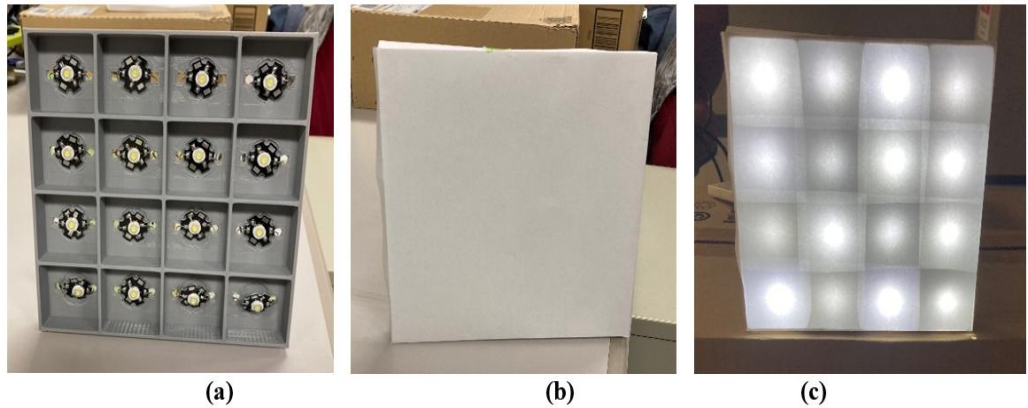

|  (a)  |  (b)  |  (c)  |

Figure 9. The prototype of the lighting panel.

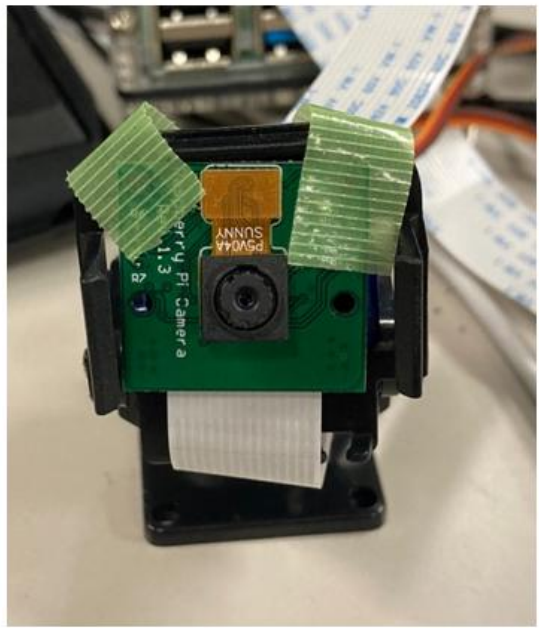

Figure 10. OV5647 Camera for Raspberry Pi.

Two sets of the marker-shaped lighting circuits are prepared, each consisting of two circuits connected in parallel. The physical setup is illustrated in Figure 11.

The marker-shaped lighting is captured using a rolling shutter camera with an exposure time of 0.01 s. The imaging distance is varied from 0.4 m to 2.0 m in increments of 0.2 m, and for each distance, 300 images are captured to evaluate the recognition rate of the AR markers. At a fixed distance of 0.6 m, the marker's angle relative to the camera is varied, with 90° defined as frontal. Angles of −70°, -45°, −20°, 0°, +20°, +45°, and +70° are tested, and the recognition rate is measured for each orientation.

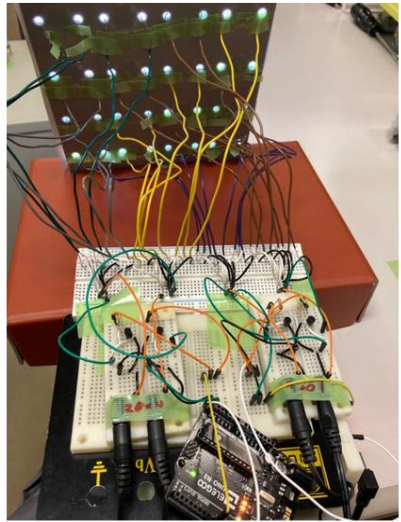

Figure 11. The physical setup with two electronic oscillator circuits.

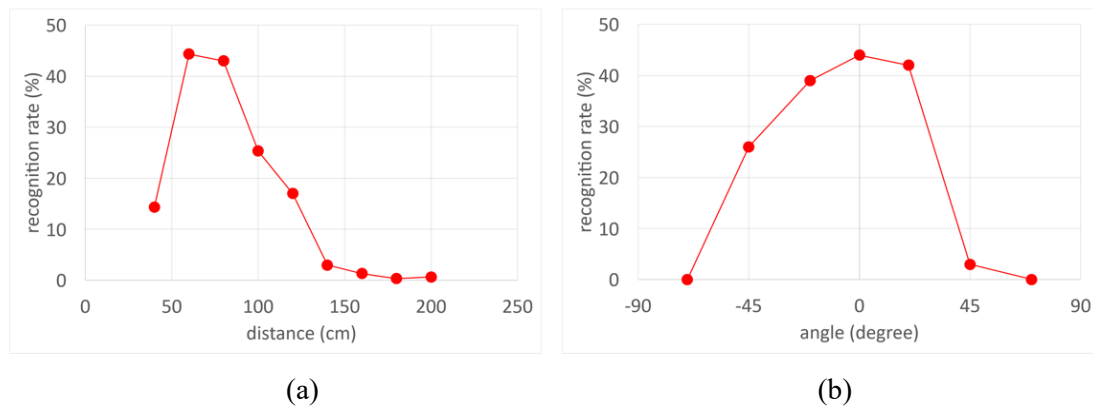

(a) (b)

Figure 12. Recognition rates of the proposed method:
(a) with respect to distance and (b) with respect to angle.

### 5.3. Results and Discussion

The recognition rate of the proposed AR marker–shaped lighting was evaluated at various distances and viewing angles. Figure 12 shows the recognition rates with respect to distance and angle.

For distance, the recognition rate was highest from 60 to 80 cm, reaching over 43%, and gradually decreased as the distance increased, which is consistent with the expected effect of image noise and light dispersion over longer distances. Although the recognition rate decreased at longer distances, this tendency does not indicate a fundamental limitation of the proposed method. Because the size of the projected marker on the image plane becomes smaller with increasing distance, the recognition performance naturally depends on the marker's physical dimensions relative to the distance. Therefore, the decrease observed in our experiments should be regarded as a relative evaluation based on the prototype size, and the method can achieve higher recognition performance at longer distances simply by enlarging the marker or using higher-resolution imaging. For viewing angle, the recognition rate remained above 40% in a range near frontal (from −20° to +20°), demonstrating robust detection under moderate angular variations. Regarding viewing

25

angles, although lower recognition rates were observed at extreme angles on one side, this asymmetry should not be interpreted as a drawback of the proposed method. Since the recognition performance should theoretically be symmetric around the frontal direction (0°), the imbalance is likely caused by environmental factors such as reflections, shadows, or slight misalignment in the measurement setup. Thus, both the distance-dependent and angle-dependent variations reflect characteristics of the experimental environment and prototype configuration, rather than inherent limitations of the LED-based AR marker approach.

These results highlight the capability of the proposed LED-based AR markers to be reliably detected by a camera. While absolute recognition rates naturally vary with distance and angle, the system shows promising performance, especially considering that the markers are perceived by humans as uniform lighting, without any visible cues. Further optimization of the marker size and image processing algorithms is expected to enhance recognition performance across a wider range of distances and angles.

Although the maximum recognition rate is limited under the present experimental conditions, this result should be interpreted as a baseline evaluation. The proposed approach is not intended for immediate deployment, and higher recognition rates are expected with larger markers or higher-resolution imaging devices.

## 6. CONCLUSIONS

In this paper, we proposed an LED-based AR marker method for information transmission and robot self-localization that can seamlessly blend into everyday environments. The recognition performance of the proposed markers was evaluated across various distances and viewing angles. The results demonstrate that the markers can be reliably detected under moderate distances and angles, despite appearing as uniform lighting to human observers.

While recognition rates naturally decrease at longer distances and more oblique angles due to image processing limitations and light dispersion, the experiments confirm that the proposed method provides robust marker detection in realistic indoor conditions. These findings indicate that, with further optimization of the marker design and improvements in image processing algorithms, the recognition performance can be enhanced, making the system suitable for practical deployment. Overall, the proposed method offers a promising approach to integrating AR markers into living spaces without affecting the human visual experience, paving the way for practical visible light communication–based self-localization systems in daily environments.

## ACKNOWLEDGEMENTS

This research was partially supported by the Research Program of Ryukoku University.

## AUTHORS

**Wataru Uemura** was born in 1977, and received B.E, M.E. and D.E. degrees from Osaka City University, in 2000, 2002, and 2005. He is an associate professor of the Faculty of Advanced Science and Technology, Ryukoku University in Shiga, Japan. He is a member of IEEE, RoboCup and others. He is a chairperson of RoboCup Japanese Regional Committee.

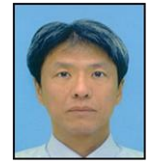

**Shogo Kawasaki** was born in 2000 and received B.E from Ryukoku University in 2023. He is interested in Visible Light Communication and Computer vision.

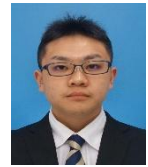